\documentclass[conference]{IEEEtran}
\IEEEoverridecommandlockouts
\usepackage[utf8]{inputenc}
\usepackage[T1]{fontenc}
\usepackage{textcomp}
\usepackage{cite}
\usepackage{amsmath,amssymb,amsfonts}
\usepackage{algorithmic}
\usepackage{graphicx}
\usepackage{textcomp}
\usepackage{xcolor}
\usepackage{subfigure}
\usepackage[shortcuts,acronym]{glossaries}
\usepackage{booktabs}
\usepackage{balance}
\usepackage[bookmarks=false]{hyperref}
\def\BibTeX{{\rm B\kern-.05em{\sc i\kern-.025em b}\kern-.08em
    T\kern-.1667em\lower.7ex\hbox{E}\kern-.125emX}}
\begin{document}
\title{Measurement-Based Modeling of Short Range Terahertz Channels and Their Capacity Analysis}
\author{\IEEEauthorblockN{Erhan Karakoca \IEEEauthorrefmark{1}\IEEEauthorrefmark{3}, 
Hasan Nayir\IEEEauthorrefmark{1}\IEEEauthorrefmark{3}, 
Güneş Karabulut Kurt\IEEEauthorrefmark{2}, 
Ali Görçin\IEEEauthorrefmark{2}}

\IEEEauthorblockA{\IEEEauthorrefmark{1} Department of Electronics and Communication Engineering, Istanbul Technical University, {\.{I}}stanbul, Turkey} 

\IEEEauthorblockA{\IEEEauthorrefmark{2} Communications and Signal Processing Research (HİSAR) Laboratory, T{\"{U}}B{\.{I}}TAK B{\.{I}}LGEM, Kocaeli, Turkey}

\IEEEauthorblockA{\IEEEauthorrefmark{3} Poly-Grames Research Center, Department of Electrical Engineering  Polytechnique Montr\'eal, Montr\'eal, Canada} 

Emails: \texttt{\{karakoca19, nayir20\}@itu.edu.tr,gunes.kurt@polymtl.ca,}\\
\texttt{{ali.gorcin@tubitak.gov.tr}}}
\maketitle
\begin{abstract}
 In this work, extensive propagation characteristics of short-range 240 to 300 GHz terahertz (THz) channels are mapped based on a measurement campaign conducted utilizing a novel, task-specific measurement system. The measurement system allows collecting measurements from different distances and orientations in a very-fine grained resolution, which is a particular issue in achieving realistic THz channel estimation. After the accurate measurement results are obtained, they are investigated in terms of channel impulse and channel frequency response. Furthermore, the fading channel amplitude histograms are modeled with the Gamma mixture model (GMM). The expectation-maximization (EM) algorithm is utilized to determine the corresponding mixture parameters. Also, to demonstrate the flexibility of the GMM, the Dirichlet process Gamma mixture model (DPGMM) is utilized in cases where the EM algorithm fails to represent histograms.
Moreover, the suitability of the GMM is evaluated utilizing Kolmogorov–Smirnov tests. Results verify that the GMMs can simulate the fading channel of micro-scale THz wireless communication in a realistic way, providing important implications regarding the achievable capacity in these channels. Finally, the average channel capacity of each link is evaluated using the probability density function of GMMs to gain deeper insight into the potential of micro-scale THz communications.
\end{abstract}
\begin{IEEEkeywords}
Terahertz, short-range communication, measurement, channel modeling, Gamma mixture model
\end{IEEEkeywords}
\section{Introduction}
The demand for wireless communications capacity has surged due to the rapid growth of connected devices. This increased demand has resulted in congestion within existing frequency bands, necessitating the exploration of new spectrums to meet data rate requirements. The terahertz (THz) spectrum emerges as a promising solution, offering wide bandwidth and enabling high data rates, while also unlocking new application areas \cite{federici2010review,tekbiyik2019terahertz}. Consequently, the THz spectrum holds significant potential for transforming wireless communications.
However, leveraging the THz spectrum for communication purposes comes with its own set of challenges. Atmospheric gases cause notable signal absorption at specific frequencies within the THz range, thereby limiting communication range. As a result, the THz band exhibits unique channel characteristics that differ from other commonly utilized sections of the wireless frequency spectrum. Additionally, different frequency windows within the THz spectrum can exhibit diverse channel characteristics \cite{akyildiz2014terahertz}.
Several measurement-based works have investigated the large-scale fading characteristics \cite{ekti2017statistical,kim2014statistical,tekbiyik2019statistical,nguyen2021large,chen2021channel,fu2020modeling,kim2016characterization,eckhardt2019measurements} to overcome these challenges and understand the intricacies of the THz band.
In~\cite{ekti2017statistical}, frequency-dependent path loss parameters of the sub-THz band are measured for line-of-sight (LOS). The single slop path loss model with shadow fading parameters is derived for the desktop channel between 300 and 320 GHz in~\cite{kim2014statistical}. The two-slope path loss model for the short-range THz propagation is given in~\cite{tekbiyik2019statistical}, which demonstrates the joint effect of the transmitter-receiver separation and the frequency on the received power. The large-scale parameters of the spatio-temporal channel for 140 GHz at the shopping mall and airport hall are investigated in \cite{nguyen2021large}. By combining ray-tracing and statistical methods, an indoor communication hybrid cluster-based channel modeling is proposed for 130 GHz - 143 GHz in \cite{chen2021channel}. Furthermore, the characterization of chip-to-chip, computer motherboard, and data center channel characterization is conducted in \cite{fu2020modeling,kim2016characterization,eckhardt2019measurements} respectively.  

The characterization of THz small-scale fading using measurements is also the subject of several studies~\cite{kim2016statistical,selimis2021initial,papasotiriou2021new,tekbiyik2021modeling,9852427,papasotiriou2023outdoor}. In \cite{kim2016statistical}, based on a two-dimensional geometrical propagation, a parametric reference model for device-to-device THz multipath fading channels is given for frequencies between 300 GHz and 320 GHz. Small-scale fading of the 140 GHz is investigated in \cite{selimis2021initial} by modeling the LOS and obstructed LOS measurements with Weibull and Nakagami-m distributions. In addition, \cite{papasotiriou2021new} investigates the suitability of the distribution for characterizing fading at 143 GHz over 4 GHz bandwidth based on data collected in a shopping mall, airport, and university. Furthermore, in \cite{tekbiyik2021modeling}, and \cite{9852427}, the utilization of Gamma mixtures is proposed with different approaches to model the wideband sub-THz channel accurately. Its suitability is demonstrated over measurements taken at 60 GHz bandwidth between 240 GHz and 300 GHz. In \cite{papasotiriou2023outdoor} also the suitability of Gaussian and Gamma mixtures to model between 140 GHz and 144 GHz is demonstrated. 

Understanding how THz communication devices perform under various real-world conditions is essential to ensure they can deliver data rates as expected. While previous works have primarily focused on THz channels with 20 GHz or less bandwidths, future applications will likely require higher bandwidths. 
To address this gap in knowledge, a specialized measurement system was designed to study the performance of micro-scale THz communication devices operating at bandwidths of up to 60 GHz.
Our setup also allows for collecting measurement data at different distances and orientations, providing valuable resources about how these devices, such as high-speed wireless connections between mobile devices, servers, and nanosensors~\cite{akyildiz2010electromagnetic} will function in actual scenarios. 
 The measurements between 240 and 300 GHz are analyzed regarding channel impulse response and frequency response. Then the small-scale fading of the sub-THz band is modeled.
Additionally, the suitability of Gamma mixtures to model the received power of measurements is investigated with the Kolmogorov-Smirnov (KS) tests. 
Furthermore, the capacity of each link is evaluated by defining the analytical probability density function (PDF) of instantaneous signal-to-noise ratios (SNRs) with Gamma mixtures.

The paper is organized as follows: Section II covers the background knowledge, including signal representation, gamma mixture model (GMM), and expectation-maximization (EM) algorithm. In Section III, the design of the measurement system and methodology are explained, along with the presentation of channel responses, channel frequency responses, and modeling of the instantaneous signal-to-noise ratio (SNR) using Gamma mixtures. Section IV evaluates the capacity analysis of the measurements based on statistical modeling. Finally, Section V concludes the study.

\section{Background}
\subsection{Signal Model}
The acquired signal can be represented as follows
    \begin{equation}
        r(t)=\mathcal{R}e\left\{\left[x_{I}(t)+j x_{Q}(t)\right] e^{j 2 \pi f_{c} t}\right\},
    \end{equation}
    where $f_c$ is carrier frequency of the transmitted signal and $j$ denotes imaginary number $\sqrt{-1}$.
    In-phase and quadrature-phase of the signal are denoted as $x_I(t)$ and $x_Q(t)$, respectively.
    The real-valued term of the complex baseband signal $r(t)$ is represented with  $\mathcal{R}e\{\cdot\}$.
    The impulse response of the complex baseband multipath channel can be given as
    \begin{equation}
        h(t)=\sum_{l=0}^{L-1} a_{l} \delta\left(t-t_{l}\right) e^{-j 2 \pi f_{c} t_{l}}.
        \label{eq:imp_resp_mult}
    \end{equation}
    where $L$ denotes the total number of multipath sources, and $\delta$ denotes the Dirac delta function. 
    Attenuation and delay coefficients for the $l^{th}$ multipath are represented by $a_l$ and $t_l$, respectively.
    Under the only LOS component condition where $L=1$, Eq. \eqref{eq:imp_resp_mult} simplifies to 
    \begin{equation}
        h(t)=a_{0} \delta\left(t-t_{0}\right) e^{-j 2 \pi f_{c} t_{0}},
    \end{equation}
    where $a_0$ is amplitude and $2\pi f_ct$ denotes considered path phase. 
    The propagation delay is defined as $t_0 =d/c$, where $d$ is the distance between transmitter and receiver, and $c$ denotes the light speed constant.
    
The measurement campaign~\cite{txmh-mz22-22} was conducted  in an anechoic room that just permits LOS propagation.
Thus, the loss factors that impact transmitted signal power can be described as antenna misalignment, hardware imperfections, and path loss combined with molecular absorption. As a result, including the influence of these attenuations on channel amplitude $a_0$, the received signal expression can be reduced to a composition of distance and frequency-dependent path loss, molecular absorption, and antenna misalignment, as given
\begin{equation}
    P_{rx}=P_{tx} - ( 10  n \log _{10}(f,d) + L_{ma}(f,d) )+ M,
    \label{path_loss}
\end{equation}
where the received power $P_{rx}$ is determined as the difference in the strength of the transmitted signal $P_{tx}$. The path loss exponent is indicated as $n$, while the distance and frequency-dependent molecule absorption is represented by $L_{ma}(f,d)$. $M$ is the random antenna gain caused by antenna misalignment.
\subsection{Gamma Mixture Model}
The Gamma distribution is flexible and can accurately represent a wide range of datasets by adjusting its parameters. Thus, it is selected as a mixture kernel to accurately represent the distributions underneath the histograms. 
The probability density function of the Gamma distribution can be given as
\begin{equation}
f(x; \alpha, \beta)=\frac{\beta^\alpha}{\Gamma(\alpha)} x^{\alpha-1} e^{-\beta x}, x \geq 0,
\label{eq:gamma_pdf}
\end{equation}
where $\alpha$ is shape and $\beta$ is the rate parameter.
Based on this, the GMM with $K$ number of compounds can be given as
\begin{equation}
p_{\gamma}\left(\mathbf{x} \mid \alpha_1, \beta_1, \omega_1, \ldots, \alpha_K, \beta_K, \omega_K\right)=\sum_{k=1}^K \omega_k f_{\gamma}\left(x;\alpha_k, \beta_k\right)
\label{eq:gamma_mixture}
\end{equation}
where $\mathbf{x}=\left\{x_{1}, \ldots, x_{n}\right\}$ is the observed values, $\omega_k$ denotes the weight for $k^{th}$ component of the mixture and their sum is one $\sum_{k=1}^{K} \omega_{k}=1$.

It has been shown that any probability density function on the positive real axis $(0, \infty)$ can be modeled using a mixture of Gamma distributions~\cite{wiper2001mixtures}. The GMM is particularly useful in approximating fading channels~\cite{atapattu2011mixture} due to its traceable cumulative distribution function and moment-generating function. Therefore, GMM is utilized for modeling the received signal power, $P_{rx}$. It is crucial to accurately extract the GMM parameters to represent the instantaneous SNR histograms analytically.
To extract GMM parameters, the EM algorithm is employed.

\subsection{Expectation-Maximization Algorithm for Gamma Mixtures}
The EM algorithm is widely used in various fields, such as statistics, machine learning, and signal processing. 
It is an iterative approach for finding the maximum likelihood estimates of parameters in statistical models containing latent variables, such as predicting mixture model parameters.

The number of mixing components must be specified to compute the EM method.
Then it updates parameter estimates through two steps until the convergence criterion 
$\left|\mathcal{L}^{[new]}-\mathcal{L}\right|<\epsilon$ holds. The $\epsilon$ denotes desired convergence value, and $\mathcal{L}$ denotes the maximum likelihood estimate of the log-likelihood function for the parameters as given
\begin{equation}
\mathcal{L}(\mathbf{x}; \alpha_{1:K}, \beta_{1:K})=\frac{1}{n} \sum_{i=1}^n \ln \left(\sum_{k=1}^{{K}} w_k p\left(x_i \mid \alpha_k, \beta_k\right)\right).
\end{equation}

The E-step estimates the membership coefficients given current parameter estimates $\theta_{1: K}=\left(\alpha_1,\beta_1 \ldots, \alpha_K, \beta_K\right)$ as 
\begin{equation}
\Omega_{i k}=\frac{\pi_k p\left(x_i \mid \theta_k\right)}{\sum_{k=1}^M \pi_k p\left(x_i \mid \theta_k\right)}.
\end{equation}
Then using the coefficients and the observed data, the mixture parameters for the GMM can be found by updating the parameter estimates in the M-step as follows~\cite{vegas2014gamma}

\begin{equation}
\pi_k^{n e w}=\frac{\sum_{i=1}^N \Omega_{i k}}{N},
\end{equation}
\begin{equation}
\mathbb{E}\left[\mathbf{x}_k\right]^{\text {new }}=\frac{\sum_{i=1}^N \Omega_{i k} x_i}{\sum_{i=1}^N \Omega_{i k}}=\alpha \beta,
\end{equation}
\begin{equation}
\operatorname{Var}\left[\mathbf{x}_k\right]^{n e w}=\frac{\sum_{i=1}^N \Omega_{i k}\left(x_i-\mathbb{E}\left[\mathbf{x}_k\right]^{n e w}\right)^2}{\sum_{i=1}^N \Omega_{i k}}=\alpha \beta^2.
\end{equation}

\subsection{Error Metric for Statistical Modeling}
As an error metric, we employed the KS test.
It is a non-parametric method used to compare the distribution of a sample data set to a reference probability distribution. It measures the maximum distance between the sample's cumulative distribution function (CDF), $F(x)$, and the reference cumulative distribution function, $G(x)$, as
\begin{equation}
    D_{\mathcal{KS}} = \max\left|F(x) - G(x)\right|
\end{equation}
where $x$ is the variable of interest and $D_{\mathcal{KS}}$ is the KS statistic. The test returns a \textit{p-value}, representing the probability that the two samples were drawn from the same distribution. 
\section{The measurement system and Methodology}
\subsection{Measurement System}
The measurement system is constructed in a fully isolated anechoic room with absorbers, and this setup can be divided into two parts; mechanical structure and electronic devices. 
The mechanical structure of the measurement system consists of a linear guide, and two sliding platforms positioned opposite each other on the guide. The rail allows these platforms to be moved and the distance between them to be adjusted as desired without losing their alignment.
The platforms also have a rotatable mechanism that enables precise angle adjustment at 1$^{\circ}$ intervals. 
The THz transceivers can be mounted to these platforms. 
Also, the angle of the transceiver is changed by taking the endpoints of the transceivers as the rotation center. This ensures that the distance between the transmitter and the receiver is maintained even if the angle changes.
Thus, it allows reliable measurements to be taken perfectly aligned and also in controlled misalignment scenarios.
\begin{figure}[ht!]
    \centering
    \subfigure[Mobile rail and the platforms placed on top of linear guide]{\includegraphics[width=0.6\linewidth]{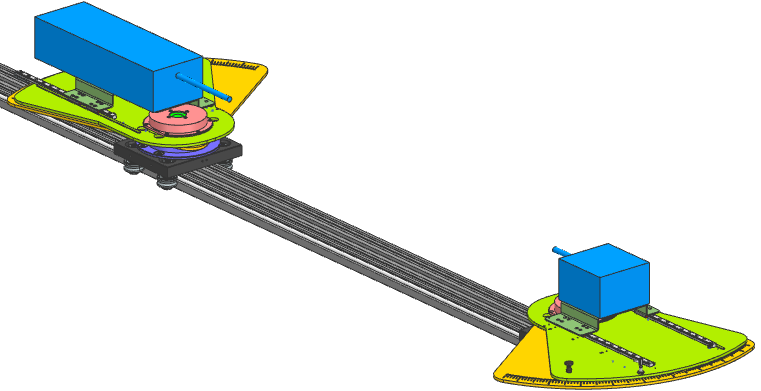}}\\
      \subfigure[Transmitter side]{\includegraphics[width=0.3\linewidth]{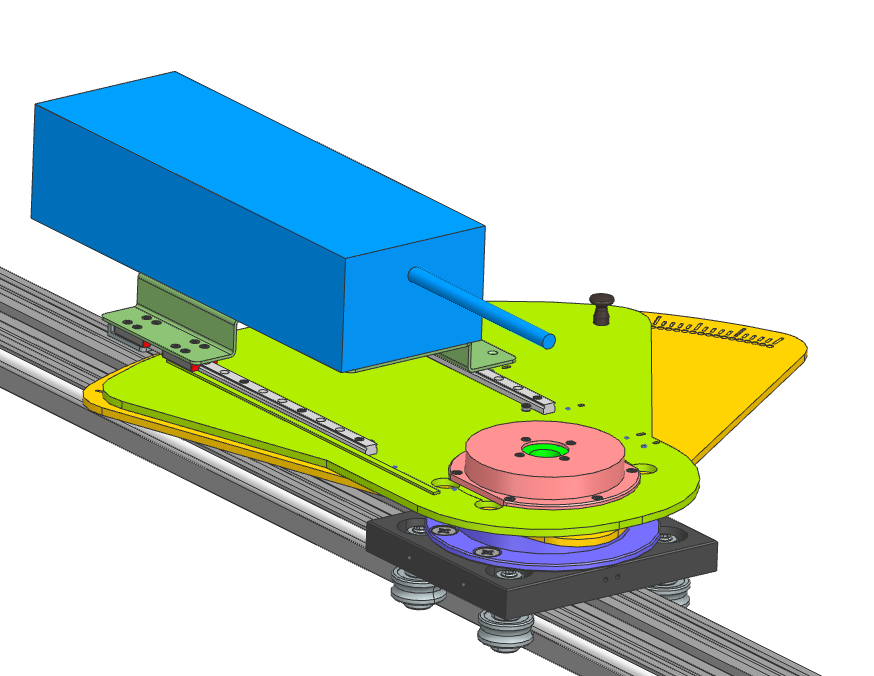}}\ \ \ \ \ \ \ \ \ \ \ \ \ 
      \subfigure[Receiver side]{\includegraphics[width=0.3\linewidth]{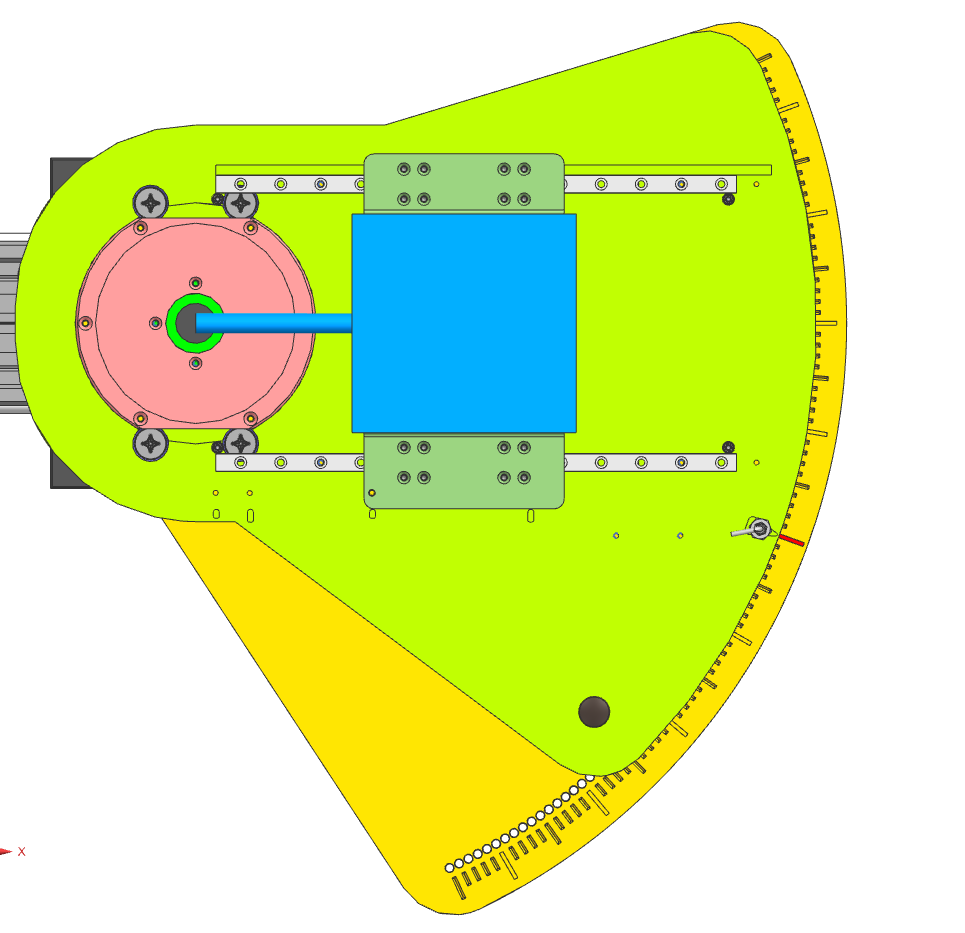}}
      \caption{3D model renderings of mechanical parts of the measurement system.}
\label{fig:mechanical_components}
\end{figure}
\begin{figure}[ht!]
    \centering
\includegraphics[width=1\linewidth]{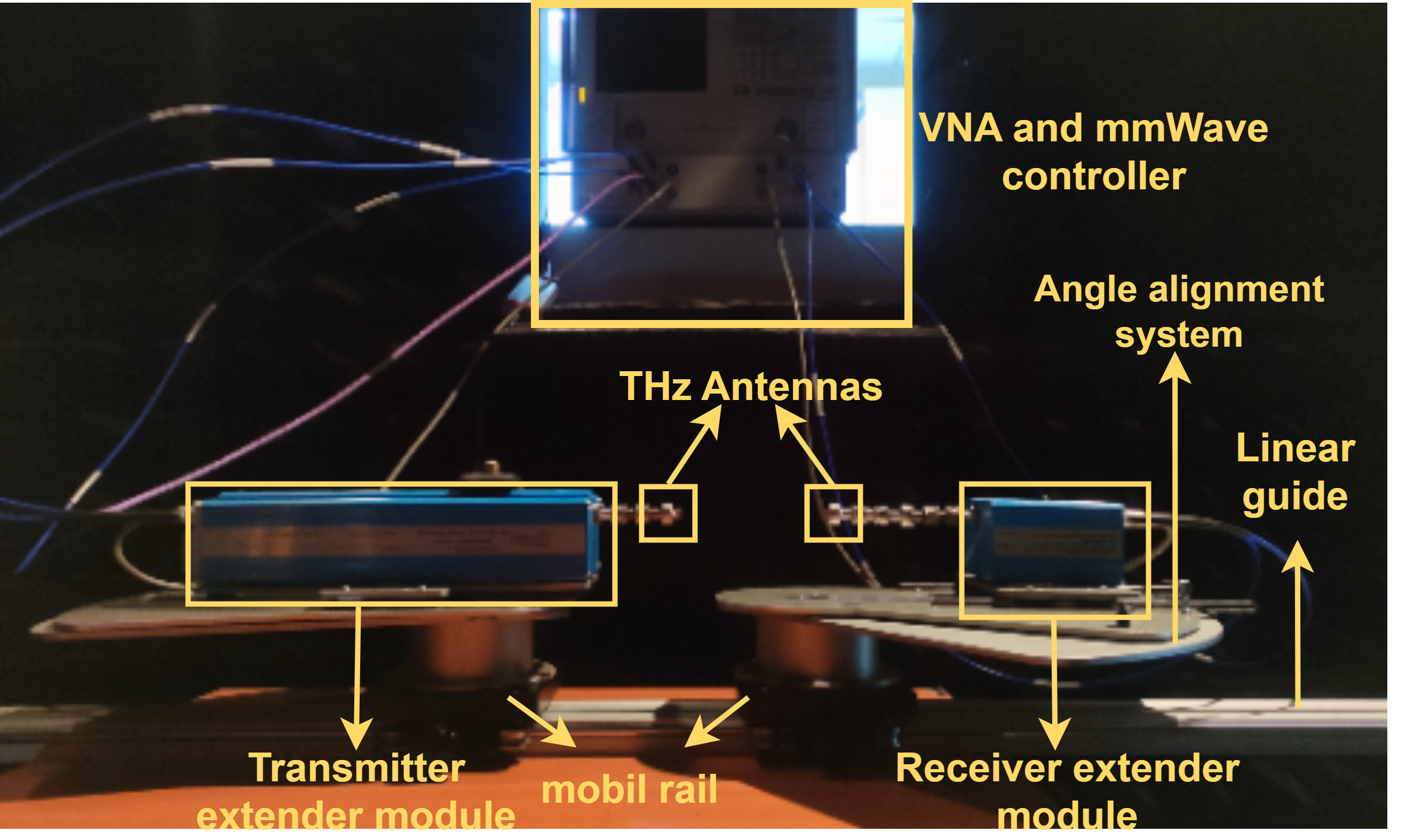}
    \caption{The measurement system at the anechoic chamber.}
    \label{fig:my_label}
\end{figure}

Electronic devices consist of three main modules; an E8361A PNA coded mm-Wave (10 MHz - 67 GHz) vector network analyzer (VNA), an N5260A coded mm-Wave controller both from Agilent, and V03VNA2-T/R-A and V03VNA2-T coded mm-Wave (220 GHz - 325 GHz) VNA extenders optimized for Agilent N5260A from Oleson Microwave Labs (OML).
VNA can analyze signals up to 67 GHz, and connecting extender modules to VNA via controller allows for analyzing signals in the sub-THz range from 220 GHz to 325 GHz. Since extenders contain 18 mixers, signals between 10 and 20 GHz can be carried up to 325 GHz. Conversely, the transmitted signal over the wireless channel in the range of 300 GHz is down-converted to the intermediate frequency (IF) between 5-300 MHz via V03VNA2-T. This way, the IF signal can be fed to the VNA as input for further analysis. Finally, the measurements can be taken as $S_{21}$ parameters via VNA.

\subsection{Measurement Methodology}
The calibration process had done before the measurements to eliminate the hardware impairments and take reliable measurements. During the calibration, the VNA initiates a known signal transmission between an end-to-end connected transmitter and the receiver extenders, then measures the losses caused by the hardware impairments between 240 GHz and 300 GHz by sweeping. These measurements are turned into the calibration data and saved as complex $S_{21}$ parameters for each point which are then employed on the experimental measurements for compensation. 
After the calibration procedure, a half-power beamwidth (HPBW) horn antenna with $24.8$ dBi gain is connected at both the transmitter and receiver.
Then, the measurements have collected at different separations for transceivers, from 20 cm to 100 cm with 10 cm increments and in different orientations with 1${^\circ}$ increments from 0${^\circ}$ to 15${^\circ}$ and with 5${^\circ}$ increments from 15${^\circ}$ to 30${^\circ}$. 
A laser-based distance meter is used to adjust the different separations accurately.
Each data collected as 4096 points of complex $S_{21}$ parameters spanning the 60 GHz spectrum between 240 and 300 GHz by sweeping. 
The configuration under consideration leads to a spectrum resolution of 14.468 MHz. Furthermore, during the experiments, the intermediate frequency (IF) bandwidth is adjusted to 100 Hz, enhancing the observed dynamic range and diminishing the noise floor.

\section{Measurement Results and Statistical Modeling}
\subsection{Channel Impulse Response}
\begin{figure}[t!]
    \centering
\includegraphics[width=0.55\linewidth]{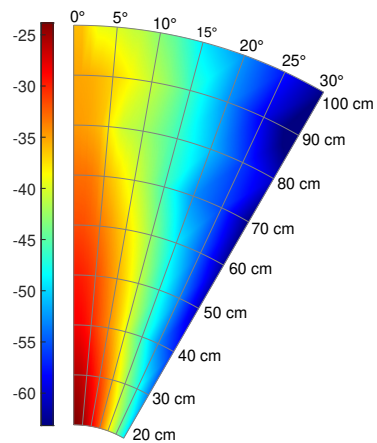}
\vspace{-15pt}
    \caption{Radial plot of channel impulse response maximum values in dB for each distance and misalignment degree.}
    \label{fig:radial_plot}
\end{figure}

The measurements are analyzed in both the time and frequency domains. The time domain analysis is done by applying the inverse fast Fourier transform (IFFT) to the measurements. 
Since all measurements are taken in the anechoic room, the considered impulse responses give only the LOS path delays. This allows for validating the measurements with adjusted distance and misalignment settings. 
\begin{figure}[!t]
    \centering
    \includegraphics[width=1\linewidth]{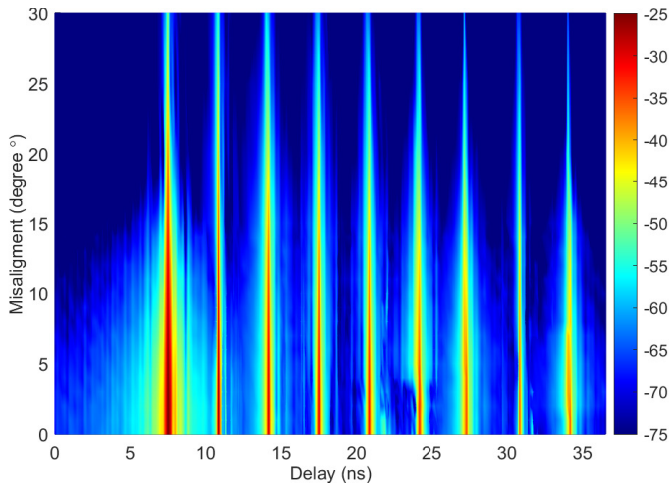}
    \caption{The plot of the channel impulse response of measurements in dB.}
    \label{fig:delay_cir}
\end{figure}
Fig. \ref{fig:delay_cir} demonstrates the combined channel impulse response for all measurement scenarios. There are signals only coming with certain delays matching to adjusted distances, which validate the measurement system for further analysis. Fig. \ref{fig:radial_plot} is produced with the maximum values of channel impulse responses for each distance and orientation to demonstrate the impact of misalignment with distance.
Also, the channel impulse response values with $5^{\circ}$ resolution for each distance are given in Table \ref{tab:cir_results}.
\begin{table}[ht!]
\centering
\caption{Received signal peak powers in dB.}
\renewcommand{\arraystretch}{1}
\setlength{\tabcolsep}{5pt}
\label{tab:cir_results}
\begin{tabular}{cccccccc}
\hline
\multicolumn{1}{l}{}        & 0${^\circ}$ & 5$^{\circ}$ & 10$^{\circ}$ & 15$^{\circ}$ & 20$^{\circ}$ & 25$^{\circ}$ & 30$^{\circ}$ \\ \hline
\multicolumn{1}{c|}{20 cm}  & -23.77      & -25.02      & -28.19       & -32.48       & -37.79       & -43.13       & -49.29       \\
\multicolumn{1}{c|}{30 cm}  & -25.86      & -27.35      & -30.95       & -35.78       & -41.71       & -47.17       & -53.70       \\
\multicolumn{1}{c|}{40 cm}  & -28.42      & -30.07      & -34.02       & -39.52       & -46.24       & -52.90       & -56.92       \\
\multicolumn{1}{c|}{50 cm}  & -29.66      & -31.38      & -35.40       & -40.74       & -47.07       & -53.25       & -60.85       \\
\multicolumn{1}{c|}{60 cm}  & -31.55      & -33.21      & -37.16       & -42.56       & -49.05       & -54.82       & -62.07       \\
\multicolumn{1}{c|}{70 cm}  & -32.92      & -35.32      & -39.82       & -45.93       & -52.73       & -55.68       & -63.20       \\
\multicolumn{1}{c|}{80 cm}  & -34.71      & -37.92      & -42.03       & -46.13       & -50.92       & -56.15       & -60.51       \\
\multicolumn{1}{c|}{90 cm}  & -35.11      & -36.84      & -40.76       & -46.27       & -52.59       & -59.08       & -66.16       \\
\multicolumn{1}{c|}{100 cm} & -35.98      & -39.92      & -43.17       & -49.14       & -52.67       & -57.57       & -63.02       \\ \hline
\end{tabular}
\end{table}
\subsection{Channel Frequency Response}
\begin{figure}[ht!]
    \centering
    \includegraphics[width=0.95\linewidth]{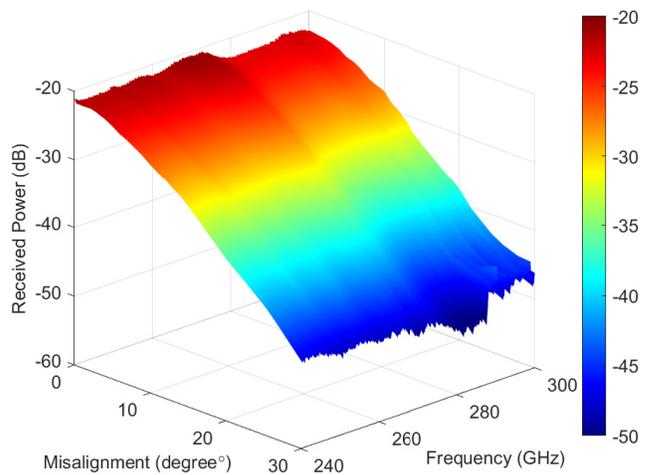}
    \caption{Channel frequency responses for 20 cm separation.}
    \label{fig:cfr}
\end{figure}

The channel frequency response between 240 and 300 GHz is investigated once the measurement system has been validated with channel impulse responses.
Fig. \ref{fig:cfr} demonstrates a channel frequency response plot at 20 cm for angles from $0^{\circ}$ to $30^{\circ}$. The distance is kept constant, but the channel frequency response is plotted for various misalignment angles for each separation of transceivers.
\begin{figure*}[!t]
  \centering
  \subfigure[20 cm $5^\circ$]{\includegraphics[width=0.329\linewidth]{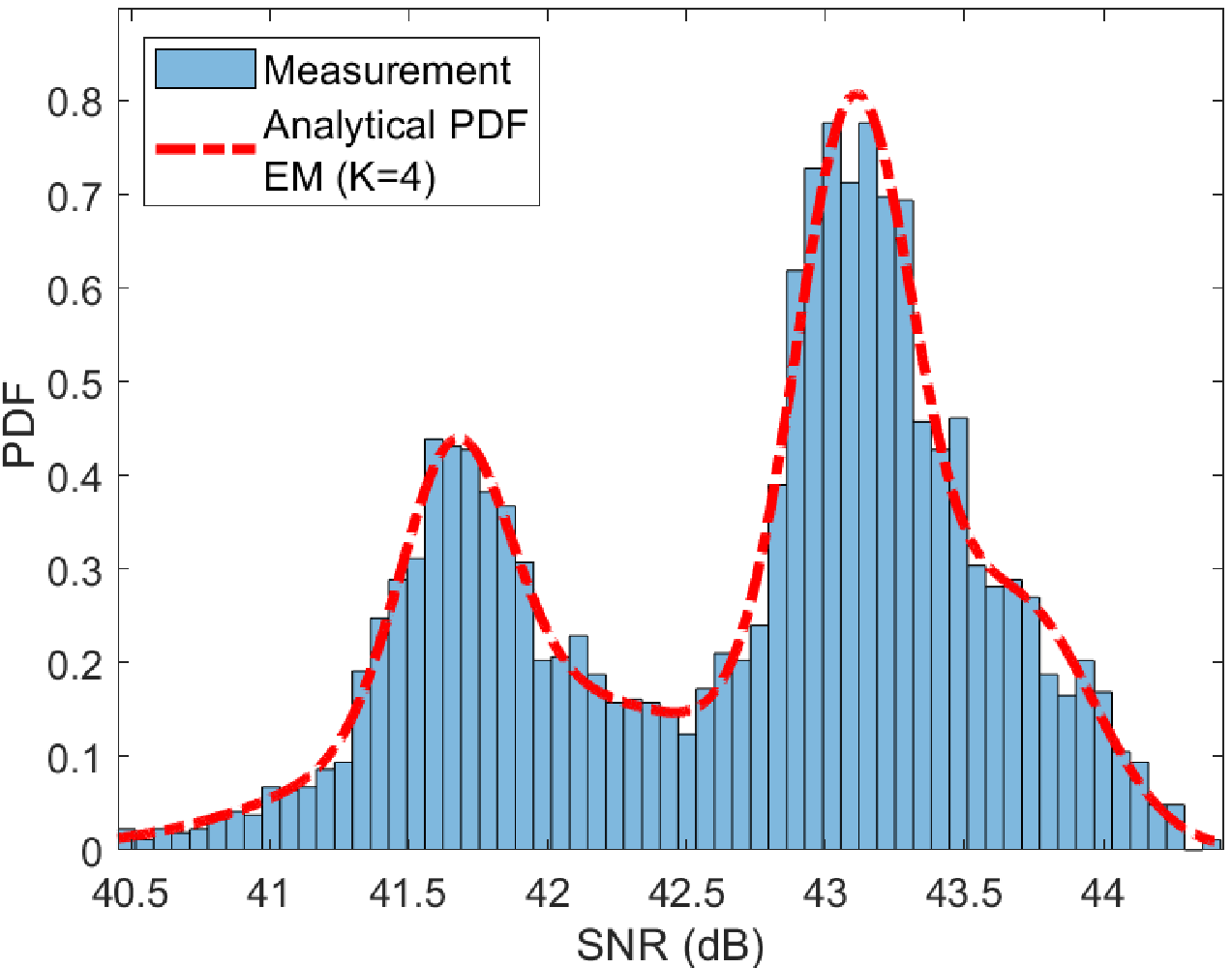}}
  \subfigure[50 cm $5^\circ$]{\includegraphics[width=0.329\linewidth]{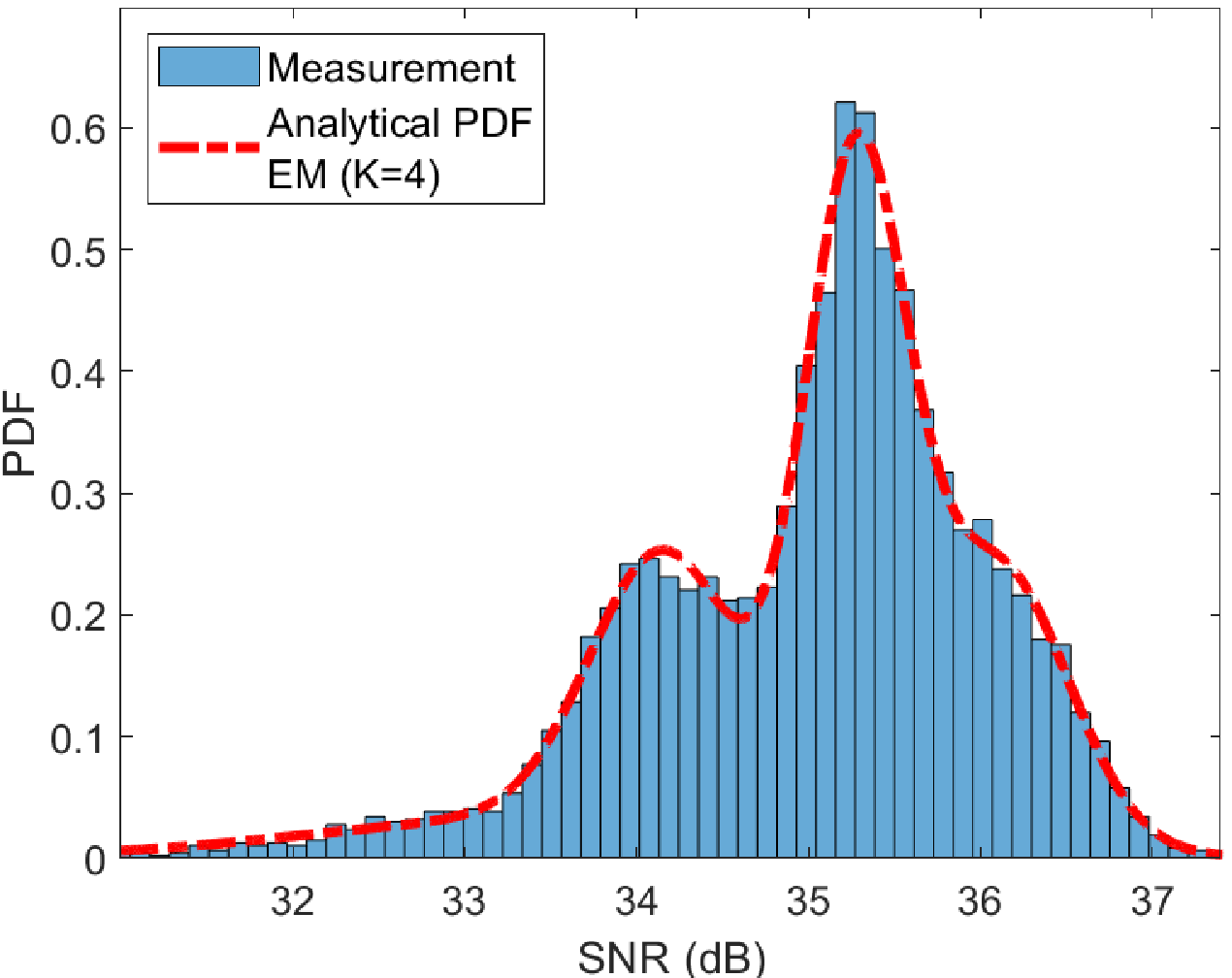}}
    \subfigure[100 cm $5^\circ$]{\includegraphics[width=0.329\linewidth]{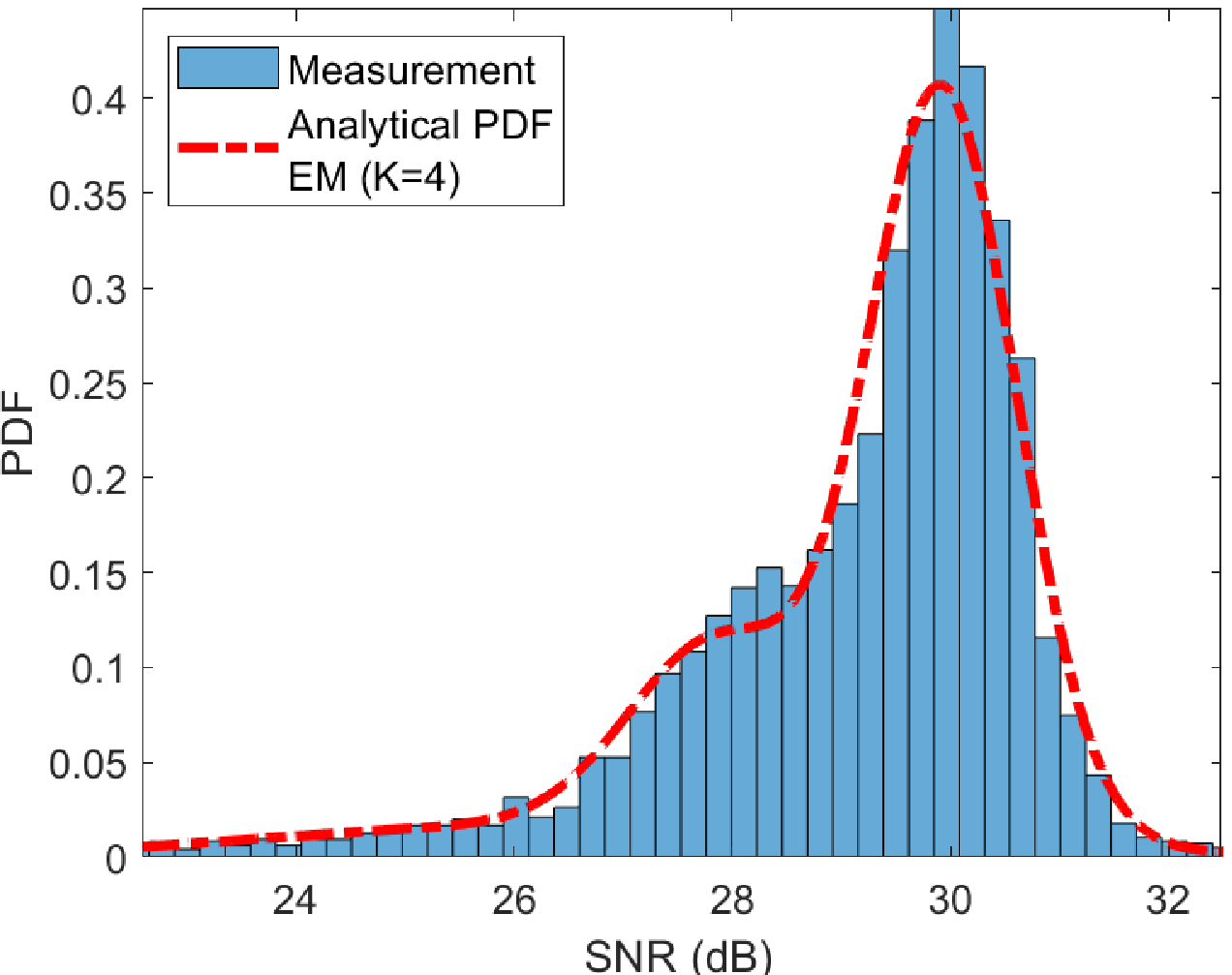}}
  \caption{The modeling of the instantaneous SNR histograms with DPGMM for 60 GHz bandwidth at different scenarios. }
  \label{fig:estimates}
  \vspace{-5pt}
  \end{figure*}
\subsection{Statistical Modeling of Instantaneous SNR}
The received power can be calculated~as
\begin{equation}
P_{r x}=\left|S_{21}\right|^2 P_{t x},
\end{equation}
where $P_{tx}$ transmitted signal power and $\left|S_{21}\right|$ is amplitude response of transmission channel obtained from measurements. Then, the instantaneous SNR can be derived from the received signal power $P_{rx}$ as follows
\begin{equation}
\gamma=\frac{P_{r x}}{B N_0},
\end{equation}
where $B$ denotes the bandwidth of the signal. 
The power spectral density, $N_0$, is assumed to be additive white Gaussian noise (AWGN).
By possessing a probability density function of the instantaneous signal-to-noise ratio (SNR), average channel capacity and outage probability can be ascertained. 
Since THz sub-bands often exhibit non-uniform and variable forms in terms of received power distribution because of their unique spectral characteristics, the instantaneous SNR values cannot be represented with a single distribution.
Also, measurements span the 60 GHz bandwidth, which can result in different fluctuations in the received power across the band owing to path loss and molecular absorption. Thus histogram of the instantaneous SNR is fitted to the given GMM, where the model parameters were estimated with the EM algorithm, and the number of mixing elements is chosen as $K=4$. 

The fitting of empirical distributions is given in Fig. \ref{fig:estimates} for three different settings.
The KS test results for the histograms are given in Table \ref{tab:ks_test}. The \textit{p-value} is set to be $0.005$ for the test.
However, since there is a total of 190 different measurement settings, we are only able to give KS test results for twenty of the measurements, which are chosen randomly.
The EM algorithm has a $90\%$ success rate in passing the KS test for 190 different settings. Nevertheless, it is not because of the Gamma kernel's incapability. It is because the EM algorithm converges to local maxima rather than global maxima. Also, convergence to local maxima is not always guaranteed, and the algorithm may become stuck at saddle points depending on initialization, resulting in incorrect parameter estimation.
Therefore we employed the proposed Dirichlet process Gamma mixture model (DPGMM)\footnote{{\url{https://github.com/erhankarakoca/DPGMM-Channel-Modelling}}} in \cite{9852427} for the cases where the EM algorithm fails to represent empirical distributions. The DPGMM was able to pass all settings that the EM algorithm failed. 
This work is conducted in view of the fact that compared to the EM algorithm, DPGMM offers the advantage of more accurate parameter estimation while having higher computational complexity.

\begin{table}[]
\renewcommand{\arraystretch}{1}
\setlength{\tabcolsep}{7.2pt}
\centering
\caption{KS-test for GMM-modeled empirical histograms}
\label{tab:ks_test}
\begin{tabular}{@{}lcclcc@{}}
\toprule
\multicolumn{1}{c}{\textbf{Setting}} & \textbf{Method} & \textbf{KS-test} & \multicolumn{1}{c}{\textbf{Setting}} & \textbf{Method} & \textbf{KS-test} \\ \midrule
20cm $5^{\circ}$                     & EM              & $\checkmark$     & 60cm $13^{\circ}$                   & EM              & $\checkmark$     \\
20cm $15^{\circ}$                    & EM              & $\checkmark$     & 60cm $20^{\circ}$                    & EM              & $\checkmark$     \\
20cm $10^{\circ}$                    & EM              & $\checkmark$     & 70cm $4^{\circ}$                     & EM              & $\checkmark$     \\
30cm $1^{\circ}$                     & EM              & $\checkmark$     & 70cm $13^{\circ}$                    & EM              & $\checkmark$     \\
30cm $5^{\circ}$                     & EM              & $\checkmark$     & 80cm $15^{\circ}$                    & EM              & $\checkmark$     \\
30cm $20^{\circ}$                    & EM              & $\checkmark$     & 80cm $25^{\circ}$                    & DPGMM           & $\checkmark$     \\
40cm $3^{\circ}$                     & EM              & $\checkmark$     & 90cm $2^{\circ}$                     & EM              & $\checkmark$     \\
40cm $25^{\circ}$                    & EM              & $\checkmark$     & 90cm $25^{\circ}$                    & DPGMM           & $\checkmark$     \\
50cm $5^{\circ}$                     & EM              & $\checkmark$     & 100cm $20^{\circ}$                   & DPGMM           & $\checkmark$     \\
50cm $30^{\circ}$                    & EM              & $\checkmark$     & 100cm $30^{\circ}$                   & DPGMM           & $\checkmark$     \\ \bottomrule
\end{tabular}
\vspace{-10pt}
\end{table}

\section{Capacity Analysis}
\begin{figure}[t!]
    \centering
    \includegraphics[width=\linewidth]{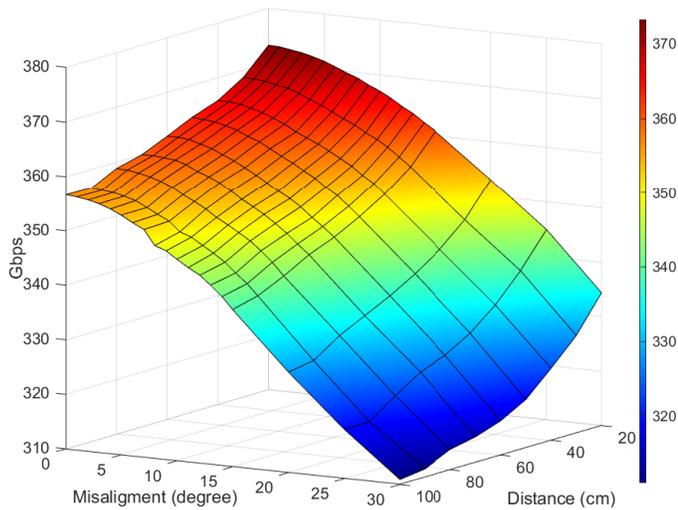}
    \caption{Capacity analysis of 240-300 GHz band for 190 different scenarios.}
    \label{fig:capacity_analysis}
    \vspace{-5pt}
\end{figure}

The capacity for the single-input single-output (SISO) channel can be evaluated using Shannon's theorem by averaging the instantaneous channel capacity over the signal-to-noise ratio (SNR) distribution as $C=\int_0^{\infty} B \log _2(1+x) f_\gamma(x) d x.$
The probability distribution function of the instantaneous SNR is defined as Gamma mixtures as given in Eq. \eqref{eq:gamma_mixture}, the capacity integral takes the form of 
\begin{equation}
    C=B \int_0^{\infty} \sum_{k=1}^K \omega_k f_{\gamma}\left(x ; \alpha_k, \beta_k\right) \log _2(1+x)  dx.
    \label{eq:capacity_int_2}
\end{equation}
 Eq. \eqref{eq:capacity_int_2} can be evaluated by replacing $\log_2(1+x)$ with Meijer G-function. Then the closed-form solution of the Eq. \eqref{eq:capacity_int_2} can be given as~\cite{atapattu2011mixture}
\begin{equation}
C=\frac{B}{\ln 2} \sum_{k=1}^K  \frac{\omega_k \beta_k^{\alpha_k}}{\Gamma(\alpha_k)} \beta_k^{-\alpha_k} G_{3,2}^{1,3}\left[\beta_k^{-1} \middle\vert\ \begin{array}{c}
1-\alpha_k, 1,1 \\
1,0
\end{array}\right],
\label{eq:closed_form_capacity}
\end{equation}
where $G$ denotes Meijer G-function. The channel capacity for the measurements can be calculated either by numerically evaluating Eq. \eqref{eq:capacity_int_2} with the analytical PDF values or by plugging the extracted mixture parameters into the Eq. \eqref{eq:closed_form_capacity}. 
We evaluated the channel capacity over the 60 GHz bandwidth for all measurement scenarios, as shown in Fig.~\ref{fig:capacity_analysis}. 
The calculations show that the maximum capacity can be achieved around 373 Gbps for the 60 GHz bandwidth at 20 cm 0$^{\circ}$ tilt, degrading as distance or misalignment increases. While a 100 cm separation causes only a 17 Gbps reduction in the capacity compared to 20 cm, 30$^{\circ}$ tilt causes approximately 40 Gpbs reduction at 20 cm.

\section{Conclusion} 
This work examines measurements collected from 190 different configurations across a 60 GHz bandwidth by analyzing the channel impulse and frequency response. The channel for future micro-scale communication systems operating within the 240 GHz-300 GHz range is studied using Gamma mixture models. The EM algorithm and DPGMM are utilized to extract the mixture parameters, and the GMM's suitability is confirmed through KS tests. By utilizing the GMM parameters, the average channel capacity is evaluated for all 190 different configurations to observe the impact of distance and misalignment on the communication systems. It is shown that antenna misalignment can cause a considerable capacity reduction in THz communication devices, even if at shorter distances. Also, the effect of the misalignment on the received power scales up with the increasing distance. Thus, in the realm of micro-scale communication systems, the incorporation of alignment algorithms and mechanisms is indispensable to establish a foundation for reliable communication between THz transmitters and receivers.
\section*{Acknowledgment}
We thank to StorAIge project that has received funding from the KDT Joint Undertaking (JU) under Grant Agreement No. 101007321. The JU receives support from the European Union’s Horizon 2020 research and innovation programme in France, Belgium, Czech Republic, Germany, Italy, Sweden, Switzerland, Türkiye, and National Authority TÜBİTAK with project ID 121N350.
\balance
\small
\bibliographystyle{IEEEtran}
\bibliography{main.bib}

\vspace{12pt}
\color{red}

\end{document}